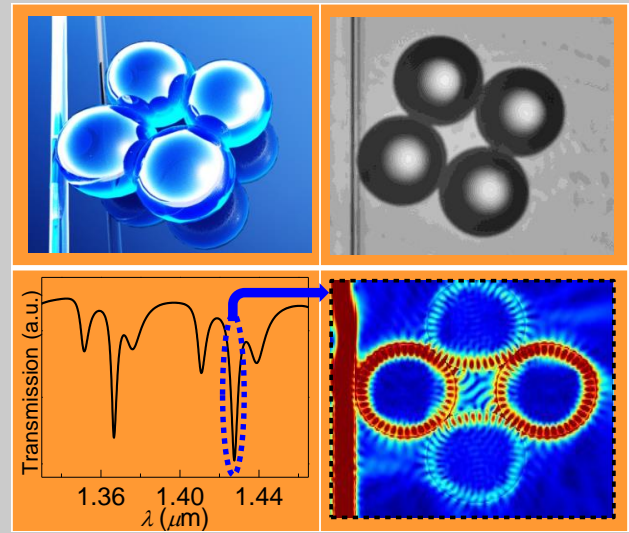

**Abstract** This work takes inspiration from chemistry where the spectral characteristics of the molecules are determined by hybridization of electronic states evolving from the individual atomic orbitals. Based on analogy between quantum mechanics and the classical electrodynamics, we sorted dielectric microspheres with almost identical positions of their whispering gallery mode (WGM) resonances. Using these microspheres as classical photonic atoms, we assembled them in a wide range of structures including linear chains and planar photonic molecules. We studied WGM hybridization effects in such structures using side coupling by tapered microfibers as well as finite difference time domain modeling. We demonstrated that the patterns of WGM spectral splitting are representative of the symmetry, number of constituting atoms and topology of the photonic molecules which in principle can be viewed as "spectral signatures" of various molecules. We also show new ways of controlling WGM coupling constants in such molecules. Excellent agreement was found between measured transmission spectra and spectral signatures of photonic molecules predicted by simulation.

# Whispering gallery mode hybridization in photonic molecules

Yangcheng Li[1,*], Farzaneh Abolmaali[1], Kenneth W. Allen[1,**], Nicholaos I. Limberopoulos[2], Augustine Urbas[3], Yury Rakovich[4,5], Alexey V. Maslov[6], and Vasily N. Astratov[1,2*]

## 1. Introduction

Artificial molecules can be synthesized from nanoscale and microscale resonant building blocks such as nanoplasmonic particles or microspherical whispering gallery resonators. The inspiration in this area comes from chemistry where molecular complexity and symmetry are central concepts for predicting hybridization of electronic states evolving from the individual atomic orbitals. In plasmonics, in recent years, all types of simple artificial molecules have been developed from metallic nanostructures including dimers [1-3], trimers [4], planar quadrumers [5], hexamers and heptamers [6]. In all these structures, the electromagnetic fields of individual particles extend away from the particles in a way reminding behavior of atomic wavefunctions. By bringing metallic particles closer to each other, a transition from isolated to collective modes of plasmonic molecules can be realized [6]. It should be noted, however, that the localized surface plasmon resonances (LSPR) associated with collective excitation of conductive electrons in the metallic nanoparticles are spectrally broadened with the typical width on the order of 20-40 nm [7]. This factor simplifies the fabrication of homogeneous plasmonic molecules where the individual LSPRs are aligned. In addition, the spectral manifestation of hybridization effects in plasmonic molecules can be described in a relatively simple way in a dipole approximation using such concepts as bonding and antibonding modes as well as dark-hot resonances [8].

In contrast, microspherical whispering gallery modes (WGMs) have much higher effective quality ($Q$) factors usually $>10^4$. The analogy between the quantum mechanics and the classical electromagnetics was noted long ago, and Stephen Arnold was the first to introduce the terminology of "photonic atoms" [9]. However, building photonic molecules from classical photonic atoms brings about an interesting problem of identity of microspherical WGM resonators required for achieving homogeneous photonic molecules and circuits. If WGM resonances are misaligned in the individual photonic atoms, their coupling and hybridization is inefficient [10-15]. On the other hand, aligning extremely narrow WGM resonances in individual microspheres represents a challenging problem. At the same time, coupled cavity structures can be fabricated without using microspheres by standard in-plane technologies, but these studies revealed the same problem related to inevitable size and shape variations of individual microdisks, rings and photonic crystal cavities with corresponding misalignment of their resonances [16-18].

---

[1] Department of Physics and Optical Science, Center for Optoelectronics and Optical Communications, University of North Carolina at Charlotte, Charlotte, NC 28223-0001, USA

[2] Air Force Research Laboratory, Sensor Directorate, Wright-Patterson AFB, OH 45433 USA

[3] Air Force Research Laboratory, Materials and Manufacturing Directorate, Wright Patterson AFB, OH 45433 USA

[4] Centro de Fisica de Materiales (CSIC-UPV-EHU) and Donostia International Physics Center (DIPC), E-20018 Donostia-San Sebastian, Spain

[5] Ikerbasque, Basque Foundation for Science, 48011 Bilbao, Spain

[6] University of Nizhny Novgorod, Nizhny Novgorod 603950, Russia

** Present address: Advanced Concepts Laboratory, Electromagnetics Division, Georgia Tech Research Institute, Atlanta, GA 30318-5712, USA

* Corresponding authors: E-mails: lycvictor@gmail.com; astratov@uncc.edu

In the first studies of simplest homogeneous molecules [19-22] this task was solved by manual sorting of few spheres with similar WGM peak positions out of hundreds of spheres with a standard size deviation of about 1%. This method made it possible for some basic studies of WGMs hybridization in bi-spheres considered as an analog of "photonic hydrogen" molecule. However, this methodology was too inefficient for developing more complicated structures and practical applications.

In parallel with building the simplest photonic molecules, an interest emerged in resonant light forces which can be used for manipulation of microspheres [23-25]. In our work, we proposed practical methods for large-scale sorting of microspheres based on using resonant optical forces [26], experimentally realized spectrally resolved optical propulsion of microspheres in evanescent field couplers [27-30], and developed the theory of these effects [31-33]. These developments in principle allow sorting large quantity of microspheres with uniquely identical resonant properties which can be used for developing new technology of heterogeneous artificial photonic molecules with the WGM resonance deviations within 0.01%.

The WGM hybridization effects in such structures are considerably more sophisticated compared to their nanoplasmonic counterparts. Such narrow WGM peaks allow observations of fine splitting effects related to symmetry and topology of the photonic molecules. Besides the purely fundamental interest, such artificial molecules can be used for engineering the density of photonic states, designing structures with the "optical supermodes" [34, 35], exploring quantum-optics analogies in photonics [36], and designing structures with unidirectional optical transport properties [37]. On a more practical level, such structures can be used for reducing the threshold of lasers, developing coupled resonator optical waveguides (CROWs) with controllable pulse delay properties, as well as sensors and filters with desired spectral characteristics.

In the present work, we study WGM hybridization in photonic molecules built by resonant microspheres with wavelength matching WGMs both theoretically and experimentally. The schematic representation of various molecules studied in planar geometry using side-coupling to WGMs is illustrated in Fig. 1. The numerical simulation is performed in 2-D case corresponding to the equatorial cross section of the 3-D structures. The modes' excitation is provided evanescently using side-coupled tapered fiber [38]. Fiber transmission spectra were presented for 3-sphere and 4-sphere chain molecules as well as 2-D molecules such as trimers, quadrumers and hexamers. The transmission spectra were investigated with three different sphere/medium index combinations.

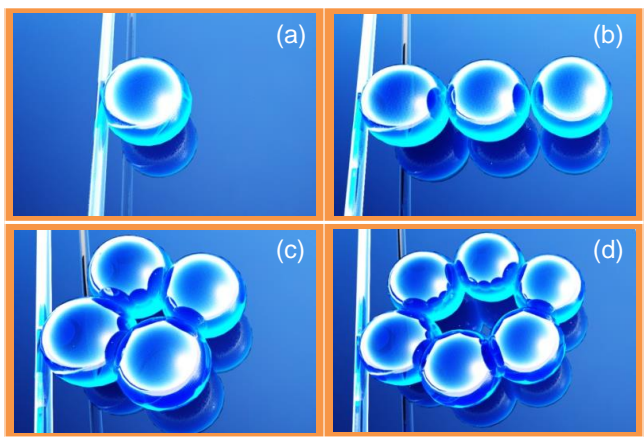

Figure 1. Illustration of the various molecular geometries with side-coupling by tapered fiber: (a) single atom, (b) linear chain, (c) planar quadrumer, and (d) hexamer.

One of the important observations of our work is that photonic molecules have certain spectral properties which are closely related to the topology and geometry of a given molecular configuration. The number of atoms determines the number of spectral components which can be split or degenerate depending on the symmetry. The spatial configuration determines the spectral positions of the split components relative to the uncoupled WGM peaks. The precise amount of splitting depends on the parameters of a given system such as the diameter and index of the microspheres and index of the surrounding medium. However, in more qualitative terms, the type of the splitting depends on the symmetry and topology of the given molecules so that it represents the number of atoms and their spatial arrangement for a given molecule. The stable spectral characteristics are represented by the total number of the split components and their spectral shifts considered in relative units. We termed the combination of these properties "spectral signature" of a given photonic molecule. Numerical methods of electromagnetic modeling used in our work do not allow us to generalize our observations and to claim that the spectral signatures are absolutely fixed characteristics. However, we show that in a certain range of variation of parameters which corresponds to a variety of practical situations, the spectral signatures are relatively stable. This means that in principle they can be used for identification of various molecules and such spectroscopic characterization can allow distinguishing between different photonic molecules. Such spectral molecular identification is of course only a distant analogue of Raman spectroscopy. Still, such identification can be made reliably.

One more property observed in our work is related to a novel way of controlling WGM coupling constants in such structures. The standard way of controlling coupling between the evanescently coupled optical components is based on controlling the nanoscale gaps separating the microresonators. Such control, however, is very difficult to implement in practice. To some extent, it can be realized using elastomeric substrates [39, 40]; however this method is difficult for practical realization.

In this work, we show that all coupling constants in the planar structures formed by microspheres can be tuned simultaneously by changing the height of the taper relative to the equatorial plane of the photonic molecule. To simplify experimental characterization, we performed characterization in an aqueous environment by working with pre-selected polystyrene microspheres with 25 $\mu$m diameter. Taking into account that the waist of the tapered microfiber has diameter of about 1 $\mu$m, changing the height of the taper in the vicinity of the equatorial plane of spheres can be achieved with micron-scale accuracy by relatively simple micromanipulation methods.

## 2. Spectral signature of photonic molecules by FDTD simulation

We used a simplified 2-D model for understanding the underlying physical properties of 3-D geometry of the optically coupled microspheres. In 3-D case, WGMs in microspheres are defined by radial ($q$), angular ($l$), and azimuthal ($m$) mode numbers [10, 19-22]. The radial number, $q$, indicates the number of WGM intensity maxima along the radial direction, the angular number, $l$, represents the number of modal wavelengths that fit into the circumference of the equatorial plane of the sphere. The azimuthal mode number, $m$, describes the field variation in the polar direction, with the number of intensity maxima along this direction being equal to $l-|m|+1$. The "fundamental" mode has $l=m$ and $q=1$. In deformed spheres, the degeneracy of azimuthal modes represented by $m$ numbers can be lifted [41]. The splitting of azimuthal modes, however, is usually observed in fluorescence measurements where all WGMs can be relatively easily excited. In contrast, in experiments with tapered microfibers the orientations of the WGMs orbits are determined by the positions where the tapered fibers touch the spherical surface. In this sense, the orientation of the modes excited in first sphere adjacent to the taper is determined by the position and orientation of the taper. The photonic molecules are assembled on a substrate, and individual WGMs in different spheres couple most efficiently in the equatorial plane parallel to the substrate. If the taper is parallel to the substrate and touches the closest microsphere in this equatorial plane, as schematically shown in Fig. 1, the coupling problem acquires some features of 2-D geometry. The spheres can be modeled as circular resonators with the diameter equal to the sphere diameter. Several circular resonators in contact position represent various configurations of photonic molecules. A tapered fiber can be modeled as a 2-D strip waveguide with 1.5 $\mu$m width and refractive index of 1.45. It was placed in close vicinity (typically 100 nm in our modeling) to one of the circular resonators for side-coupling. The WGMs in such 2-D geometry are described by radial and angular numbers [42]. This model is also applicable to cylindrical resonators such as semiconductor micropillars [43, 44].

The numerical simulation was performed by finite-difference time-domain (FDTD) method [45] with commercial software by Lumerical [46]. A Gaussian modulated pulse of ~10 femtoseconds width was launched into the waveguide from the bottom upwards, as illustrated in Fig. 2. The electric vector of electromagnetic waves was linearly polarized in the plane. In 3-D case, this situation has a close analogy with excitation of TM polarized WGMs in the equatorial plane of spheres. The transmission through the fiber was calculated as a function of time, and the broadband transmission spectra were obtained using Fourier transform [47, 48]. Due to WGMs' nature of trapping light, the simulation involving microresonators needs sufficient time to allow multiple pulse circulations and pronounced decay of the coupled light. We chose the simulation shut-off criteria as when the electric field in the computational space is less than $10^{-5}$ of the input field and found the results converged well.

Four configurations of photonic molecule were studied, including linear chains of 3 circular resonators (Fig. 2(b1)) and 4 circular resonators (Fig. 2(c1)), planar quadrumers formed by 4 ciricular resonators (Fig. 2(d1)) and hexamers formed by 6 circular resonators (Fig. 2(e1)). In order to find the spectral signature commonly seen in each configuration, the simulations were performed for three different combinations of the structural parameters. The refractive indices of the medium and the microspheres ($n$) as well as their diameters ($D$) are shown respectively for these three cases in the top row of Fig. 2. Appropriate diameters were chosen for each case to ensure WGMs of first radial order ($q=1$) are well pronounced while second order modes are suppressed. This approach enables unambiguous observation of supermodes in the fiber transmission spectrum that arises from inter-resonator coupling of WGMs with the same radial and angular modal numbers.

The transmission spectra of side-coupling to single circular resonators are presented in Figs. 2(a2-a4) respectively for these three cases. The spectra show periodic dips due to coupling of light to WGMs in circular cavities. The WGMs of a single resonator are referred to as uncoupled modes here to be distinguished from supermodes in photonic molecules. Simulation results for molecule configurations in Figs. 2(b1-e1) are presented in their corresponding rows. Strong coupling and mode splitting are observed in every case. Blue vertical line indicates the wavelength of the uncoupled mode, which helps to compare the relative positions of the split supermodes (in the case of hexamer molecule the position of the uncoupled mode is shifted compared to other structures). Red ellipse marks a set of supermodes that arises from inter-resonator coupling of one WGM, representing a reproducible unit that periodically appears in the spectrum separated by the free spectral range. As one can see, the splitting patterns of supermodes are quite similar even though the diameter of circular resonators is varied from 7 $\mu$m to 25 $\mu$m, their index was varied from 1.59 to 1.9, and the medium was changed from air to water. As an example, in Figs. 2(b2-b4) the uncoupled WGMs from single resonator are split into three supermodes. These modes are almost equally separated, and the central one has much higher magnitude, as interpreted by the depth of the dip in the transmission spectrum. A slight red shift of the central mode compared to the uncoupled mode is also noticeable in all cases. As a summary of the splitting patterns commonly seen in all three cases, we plotted supermodes' positions relative to the position of uncoupled mode in Fig. 2(b5). Relative distance from each supermode to the uncoupled one is shown as $L_N$, where number $N$ represents the order in which a given resonance appears in the transmission spectrum from left to right. We call this stable property represented by the total number of split supermodes and their relative positions to the uncoupled mode as the spectral signature specific to such photonic molecule.

Spectral signatures summarized for each photonic molecule are presented in Figs. 2(b5-e5). In the case of linear chain molecules the total number of supermodes is equal to the number of constituting atoms. For a linear chain of 3 circular resonators in Figs.2 (b2-b4), the position of the central supermode is slightly red shifted ($L_2$) relative to the uncoupled mode [49]. The splitting is almost symmetric with equal separation between adjacent supermodes. These observations are in agreement with Bloch modes formed in a coupled linear structure [14]. However, for 2-D molecules with non-straight geometry the symmetric splitting is no longer present. Transmission spectra in Figs. 2(d2-d4) show uncoupled mode splitting into three supermodes when a planar quadrumer is formed by 4 circular resonators. In comparison to a 3-sphere chain which also yields three supermodes, the modes in the square molecule (Fig. 2(d2-d4)) have wider and uneven splitting ($L_1+L_2$ and $L_3-L_2$) and larger red shift ($L_2$) for a central supermode. The ratio $(L_1+L_2)/(L_3-L_2)$ is found to vary in structures with different index and size of the constituent atoms. The combination of these properties can in principle be used for distinguishing such quadrumer's molecules from linear chains based on the spectral analysis. Similar properties can be noticed when comparing hexamers formed by 6 circular resonators to a 4-sphere chain. While this ring molecule gives the same number of supermodes (four) as the chain, the separation between the two central modes ($L_2+L_3$) is substantially larger than ($L_1-L_2$) and ($L_4-L_3$), thus enabling distinction of the hexamer molecule from other molecules considered in Fig. 2. It should be noted that the degeneracy may not be completely lifted due to symmetry of such molecules, resulting in less observed modes than the total number of atoms $N$. However that is also a distinct property which can be used for identification of given molecules. Each photonic molecule with a particular configuration and symmetry has compared to the uncoupled WGMs, as seen in Figs. 2(c4,d4,e2). The $Q$-factor increase may relate to the symmetry of the molecule configuration that was reported before [50, 51]. Hence, the configuration could be designed for potential applications such as high order filters and multi-wavelength sensors [52]. It should be noted, however, that the optical transport and coupling process is more complicated in large 2-D molecules, which provide multiple propagation paths and coupling possibilities. As a result, the spectral signatures for large molecules can be rather complicated and, in addition, their spectral appearance can vary in different spectral ranges. The spectral manifestation and shape of the resonances depends on the phase relationships in the system which can vary along the spectrum. These properties still require further investigation.

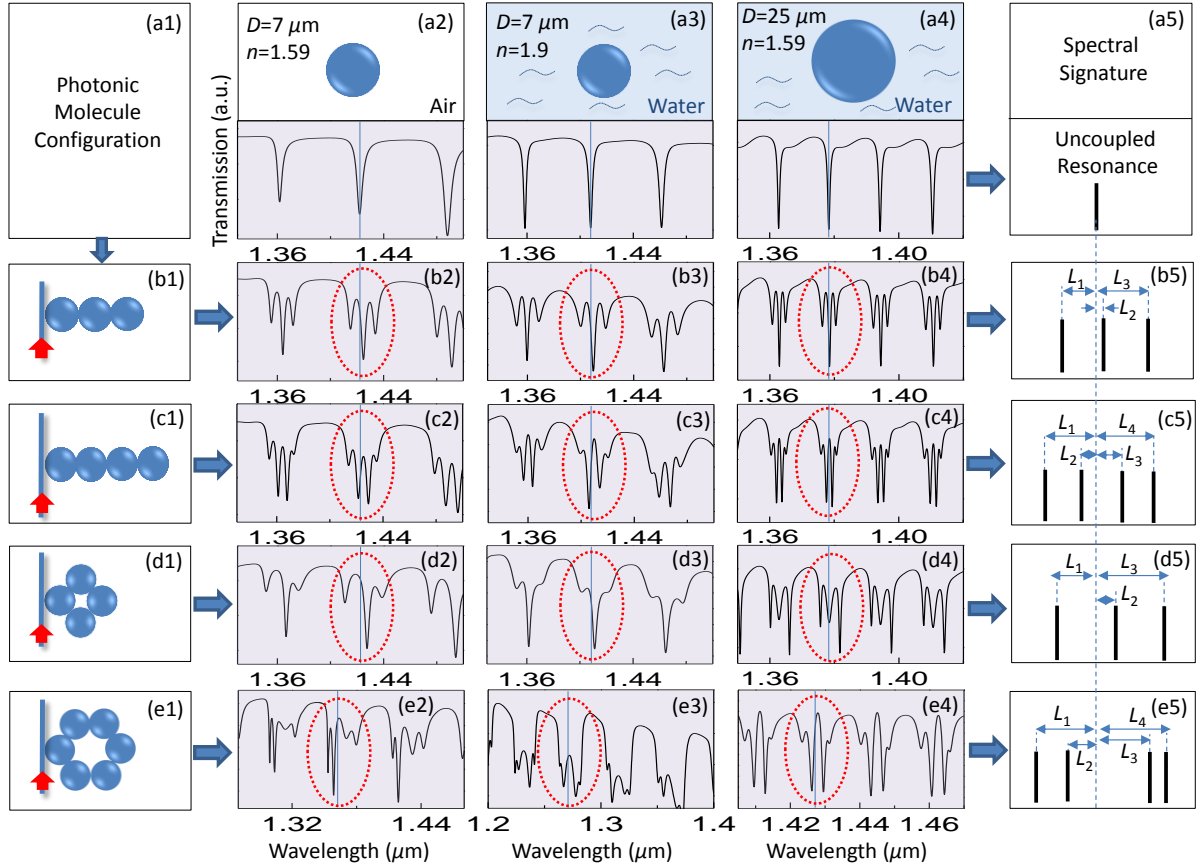

Figure 2. (a1) Molecular configurations including: linear chain of (b1) 3 circular resonators and (c1) 4 circular resonators, and planar (d1) quadrumers, and (e1) hexamers. Simulated transmission spectra of (a2-a4) single resonators and various photonic molecules (b2-b4, c2-c4, d2-d4, and e2-e4). In these spectra, the column (a2-e2) corresponds to $D=7$ $\mu$m with $n=1.59$ in air, the column (a3-e3) corresponds to $D=7$ $\mu$m with $n=1.9$ in water and the column (a4-e4) corresponds to $D=25$ $\mu$m with $n=1.59$ in water, respectively. (a5-e5) Spectral signatures for corresponding photonic molecules.

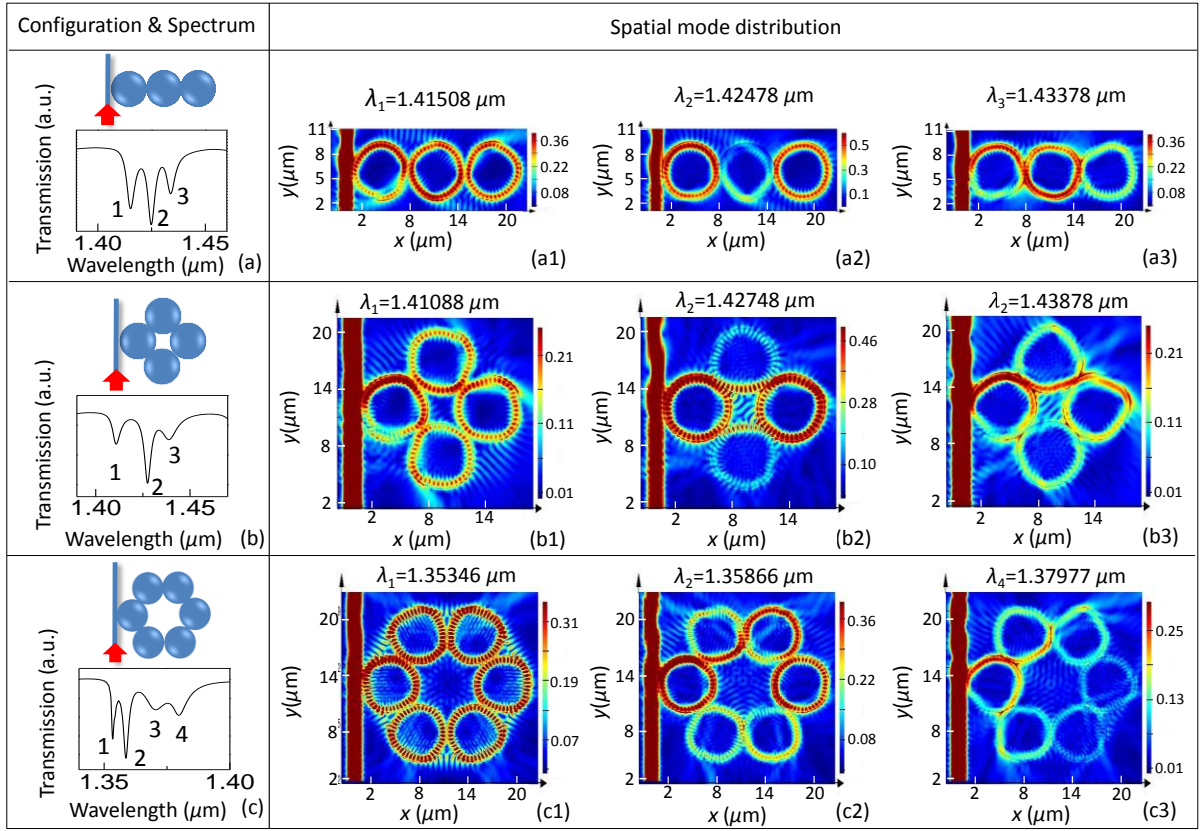

Figure 3. Various photonic molecules: (a) linear chain formed by 3 circular resonators and planar (b) quadrumers, and (c) hexamers. The split WGM components are numbered in (a-c). Simulated EM field map at three different supermode eigen wavelengths are presented for each molecule in (a1-a3), (b1-b3), and (c1-c3), respectively.

unique resonant properties that give rise to its distinct spectral signature. The spectral signature of quadrumers formed by 4 circular resonators (Fig. 2(d5)) is very different from that of 4-sphere chains (Fig. 2(c5)) and 3-sphere chains (Fig. 2(b5)). Therefore we are able to identify the molecule configuration based on its spectrum, giving us the ability to potentially utilize such signatures for geometry or position sensing. We also observed that some of the coupled supermodes have much higher $Q$-factors

## 3. Spatial distribution of supermodes in photonic molecules

To better understand the coupling and transport properties of photonic molecules, we mapped the spatial distribution of electric field (E-field) for each supermode, as presented in Fig. 3. The simulations were performed for circular resonators with $D = 7$ $\mu$m and $n = 1.59$ in air, the same as in Figs. 2(a2-e2). Photonic molecule configurations and their corresponding transmission spectra are shown in Figs. 3(a-c), where vertical lines indicate the position of uncoupled WGMs. Spatial E-field distribution was obtained by launching contiunous wave (CW) source at each supermode eigenwavelength.

Two interesting phenomena can be observed in the E-field map. Based on analogy with the well known case of two identical strongly coupled circular resonators, the shortest and longest wavelength components can be called antibonding and bonding modes, respectively [22, 51]. For shortest wavelength antibonding mode (Fig. 3(a1-c1)) the electric field appears to be distributed uniformly among all constituent atoms, which is not seen in other modes. The E-field is reduced at the points where the circular resonators touch which means that the coupling at these points is weakened. However, such antibonding modes seem to possess an additional coupling mechanism due to "leaky" modes propagating long distances along the peripheral area of the molecule. This phenomenon is especially noticeable in the hexamer molecule. As shown in Fig. 3(c1), the E-field standing waves present in the medium connecting adjacent resonators, constituting a large outer ring. These phenomena are related to the antibonding nature of such modes. The E-field is "repulsive" between two neighboring resonators due to their phase mismatch at the contact point, thus the light tunneling is less efficient. In the meanwhile, light scattered into the medium that is incident on the adjacent sphere at a grazing angle can be coupled outside the contact region, which may contribute significantly to the supermode coupling. As a result, relatively large fraction of the mode is present in the external medium leading to a smaller effective index for such antibonding mode, which results in its shortest wavelength position among all split components. As seen in Fig. 2, this antibonding mode in hexamer molecule (Fig. 3(c1)) has much higher $Q$-factor compared to the uncoupled mode ($Q$ increases from ~300 to ~1000). Similar $Q$-factor increase for some of the supermodes of symmetric photonic molecules was previously observed and studied as a function of the gaps between coupled resonators [50].

For bonding mode, in contrast, the E-field is better overlapped at the contact region adjacent to the point where two resonators touch due to improved phase matching. This factor provides more efficient evanescent side-coupling for WGMs in neighboring cavities. As one can see in Figs. 3(a3-c3), the intensity distribution indicates that light is more likely to tunnel to the adjacent resonator than to stay in the original one and circulate. Therefore, the light propagates through the entire molecule with a snaky path, tunneling at the contact point. The hexamer molecule is likely to operate in the over-coupled regime, due to the fact that bonding mode (Fig. 3(c3)) is broader and shallower and has lower $Q$-factor compared to the antibonding mode illustrated in Fig. 3(c1).

For photonic molecule supermodes whose eigenwavelengths are close to the WGMs wavelengths of the constituting circular cavities, the E-field patterns inside individual resonators appear more perfect circular-shaped, as seen in Fig. 3(a2-c2), similar to the WGMs E-field map inside a single resonator. It is due to the fact that the phase conditions inside individual atoms are closer to their uncoupled WGMs which have perfect circular symmetry. If the supermode eigenwavelength is far away from that of the uncoupled mode, large phase mismatch will create the E-field maps inside such resonators which are far from perfect circles (left resonator in Figs. 3(a1,b1)) or distorted uncircular shapes (Figs. 3(b3,c3)). It appears that light is forced to choose an optical path other than the perfect circle in order to maintain the phase matching condition. Previously, a similar effect was observed in size mismatched bi-spheres [12]. Another noticeable point in Figs. 3(a2) is that the $E$-field is concentrated in two side resonators, while the central one is almost dark. For linear chain this can be explained by the Bloch modes formation in the coupled molecule. Ref. [14] proved by calculations and demonstrated experimentally that among the three split modes in a coupled 3-sphere chain, the center mode is dominant for the first and the third resonators while the split-off modes showing more intensities in the cental resonator. For 2-D molecules with a configuration different from the linear chain the coupling is more complicated, however, similar effects can be observed in Fig. 3(b2). Enhanced E-field is also observed in the central region enclosed by four resonators due to interference. The E-field enhancement seen at the center of quadrumer in Fig. 3(b2) and along the peripheral area of hexamer in Fig. 3(c1) is advantageous in sensing applications due to strong light-particle interaction in these regions [52], not limited to the surface of the resonator like in typical WGM-based sensors.

## 4. Experimental apparatus and procedure

To study the spectral properties of photonic molecule supermodes, we experimentally assembled various molecular configurations with pre-sorted polystyrene microspheres with nominal diameter of 25 $\mu$m and similar resonant positions of WGMs. The positions of the WGM resonances were determined using transmission spectra through side-coupled tapered fiber in aqueous environment. Liquid environment enables particle movement and is critical for developing biomedical applications, thus most of WGMs based sensors are characterized in an aqueous medium [53, 54]. Due to reduced index contrast in water (~1.57/1.33 at $\lambda$~1.3 $\mu$m), the polystyrene micrsospheres should be sufficiently large ($D$>15 $\mu$m) for achieving $Q$>>100 [27-30]. Large size of microspheres was convenient for aligning the tapered microfiber with the equatorial plane of microspheres. To the best of our knowledge the present work is the first experimental demonstration of coupled photonic molecule in aqueous environment.

The selection of resonant microspheres was achieved by spectroscopic characterization and comparison. The experimental platform for sphere selection is depicted in

Fig. 4(a). A piece of single mode optical fiber (Corning SMF-28e+) transmitted through the sidewall of a plexiglass frame was wet etched in hydrofluoric acid to achieve a waist diameter of ~1.5 $\mu$m with several millimeters in length [27-30, 38]. The fiber was connected to a broadband unpolarized white light source (AQ4305; Yokogawa Corp. of America, Newnan, GA, USA) and an optical spectral analyzer (AQ6370C-10; Yokogawa Corp. of America) for transmission measurements. The frame was sealed with a microscope slide at the bottom and then filled with distilled water. The spheres used in our experiments are polystyrene microspheres (Duke Standards 4000 Series, Thermo Fisher Scientific, Fremont, CA, USA) with refractive index of 1.57 at 1.3 $\mu$m wavelength where transmission spectra were measured. Monodisperse polystyrene microspheres were chosen due to their good size uniformity of 1-2% standard deviation. Because of the reduction of the index contrast in water, the spheres with large diameter are desirable to preserve sufficiently high $Q$-factor [27-30, 38].

We found that 25 $\mu$m diameter spheres are a good candidate. They show well-pronounced first-order ($q=1$) WGMs dips in water ($TM_{q=1}^{l}$, $TE_{q=1}^{l}$), where $l$ is the angular number representing the number of modal wavelengths that fit into the circumference of the equatorial plane of the sphere. It is expected that WGMs are excited in the plane containing the microfiber and passing through the sphere center. For each position of the taper several azimuthal modes with close $m$-numbers can be excited simultaneously [55]. These modes are degenerate in a perfect sphere, however they are split in energy in real physical microspheres due to their slight (<1%) uncontrollable ellipticity [41]. The observed dips in the fiber-transmission spectra can be broadened by the partial spectral overlap of these modes. This can lead to a reduction in the measured $Q$-factors. As seen in Fig. 4(b), WGM $Q$-factor ~$10^3$ is observed while second-order modes are not present, which enables identification of the $l$-numbers. In our experiments, these numbers were determined by fitting the positions of resonant dips in the long-range (>200 nm) fiber-transmission spectra using the Mie scattering formalism [38, 55]. The $l$-numbers for TE and TM polarized WGMs are indicated in Fig. 4(b).

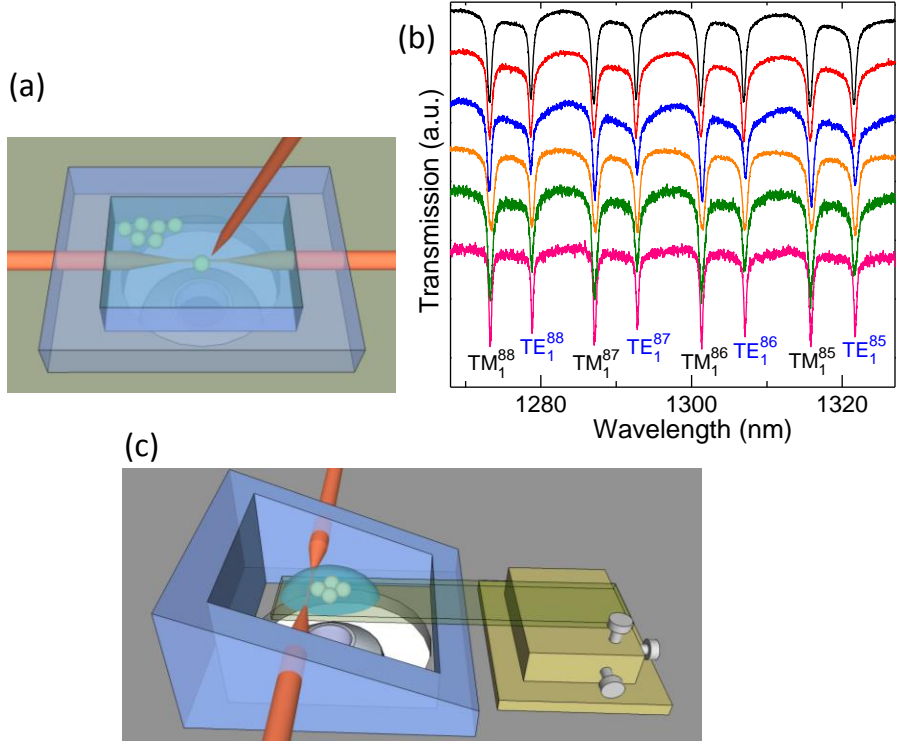

Figure 4. (a) Experimental platform for selection of size-matched microspheres by spectroscopic characterization and comparison. (b) Transmission spectra of six selected size-matched spheres with WGM resonance wavelength deviation within 0.05%. (c) Experimental set-up for side-coupling between assembled photonic molecule and tapered fiber.

The large sphere size also simplifies manipulation with individual spheres. As shown in Fig. 4(a), a tapered fiber stick was used to pick up the sphere from the bottom and attach it to the tapered region of the microfiber. For each sphere broadband transmission spectrum was recorded and compared with the spectra of other spheres. Spheres with identical positions of WGM resonances were collected in a designated area. Such sorted spheres can be used for building photonic molecules from homogeneous atoms. Fig.

4(b) shows transmission spectra of six sorted spheres in a broad wavelength range, demonstrating extraordinary high size uniformity with a standard deviation within 0.05%.

These pre-sorted spheres with almost identical resonant properties were assembled into certain molecular configurations with micromanipulation on a glass slide, which was attached to a 3-D micromanipulation stage. In our experiments we used a thin (few microns) layer of epoxy at the surface of glass plate to fix the spheres in a position. To provide access to the taper as a result of horizontal insertion of the slide with microspheres, a wedge-shaped plexiglass frame was fabricated, as sketched in Fig. 4(c). Droplets of distilled water were deposited on the glass plate to create an aqueous environment for the spheres. Then the position of the assembled molecule (quadrumer is shown in Fig. 4(c)) was adjusted by the 3-D stage to get in contact with the tapered region of the fiber for transmission measurement.

## 5. Experimental results and discussion

We began our experimental studies with the bi-sphere photonic molecule due to its simplicity. With the procedure presented in Section 4, two polystyrene microspheres of identical eigenwavelengths were selected and assembled on the glass plate in touching position. The spheres were immersed in water and brought to be in contact with the tapered fiber oriented perpendicular to the bi-sphere axis, as schematically sketched in Fig. 5(a). The position where the taper touches the sphere was carefully adjusted relative to the equatorial plane while fiber transmission spectra were continuously recorded.

When the contact position of the fiber was close to the equatorial plane of the bi-sphere molecule, short-wavelength antibonding mode and long-wavelength bonding mode were well pronounced and normal mode splitting could be measured, as seen in Fig. 5(b). This splitting was found to be highly sensitive to the position of the taper relative to the spheres' equatorial plane. We repeatedly moved the slide with microspheres up and down and kept the fiber in contact with the bi-sphere while monitoring the transmission spectra in the meantime. Several examples of fiber transmission spectra measured for different height differences $h$ between the fiber and the equatorial plane are presented in Fig. 5(b), showing the decrease of splitting amount with increasing $h$. It is well known that the amplitude of normal mode splitting is a representation of coupling strength, and the coupling constant can be derived from the splitting amount ($\Delta\lambda$) as $\kappa = 1/\sqrt{3} \times \Delta\lambda/\lambda$ for bi-sphere [14, 22]. The calculated values of $\kappa$ as large as $1.6 \times 10^{-3}$ in the equatorial plane are indicated on the corresponding spectra in Fig. 5(b).

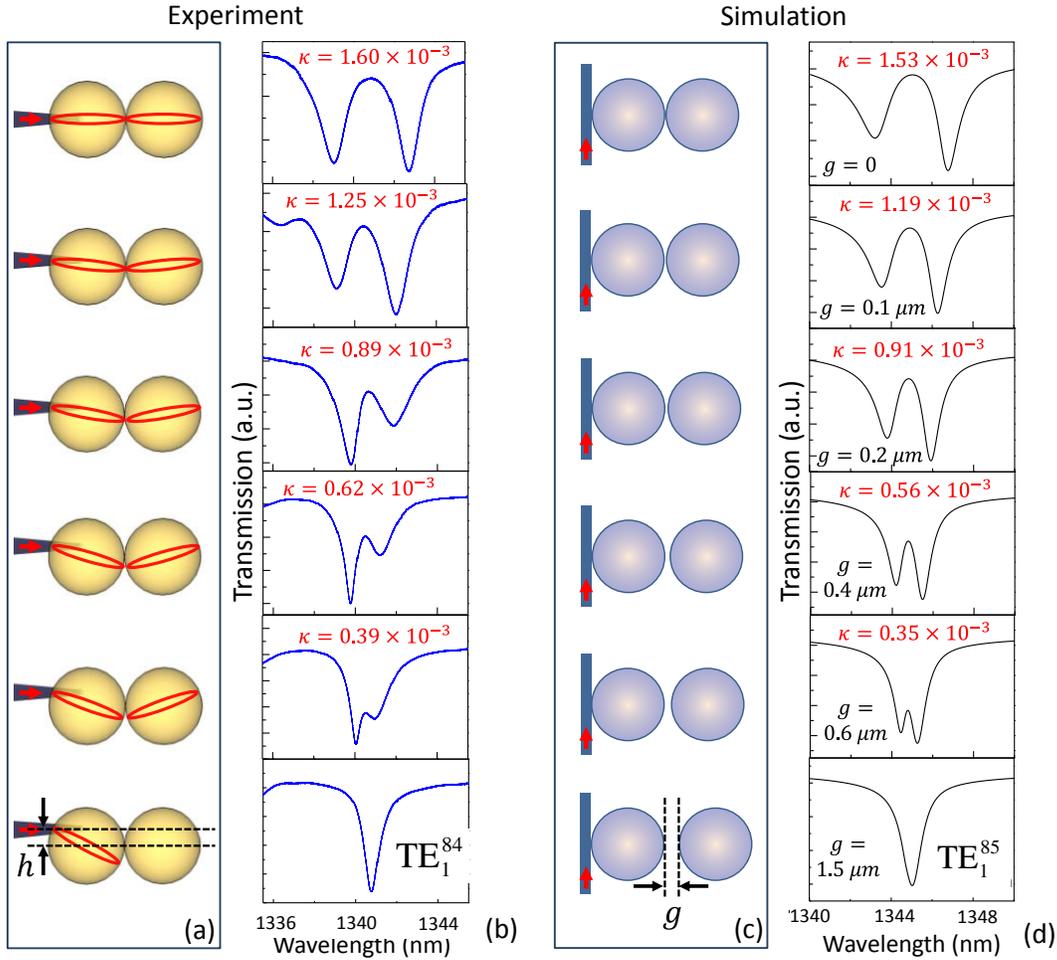

Figure 5. (a) Increase of the tapered fiber offset $h$ from the equatorial plane of bi-sphere with (b) corresponding experimental transmission spectra. (c) Increase of inter-sphere gap $g$ in bi-sphere with (d) corresponding transmission spectra in simulation.

The reduction of the coupling strength with the increase of $h$ is explained by geometrical factors. Since the taper touches the surface of microsphere tangentially, its spatial orientation and position determines the deviation of the WGM orbit from the equatorial plane. The exact description of such modes is rather difficult and requires a single WGM with the plane of the orbit tilted relative to the plane of the substrate. In this simplified model, we neglect with the substrate influence on WGMs. Due to large sphere diameters and small index contrast, the effective length of the "coupler" region, the area near the point where the spheres touch, is sufficiently long, so that the momentum conservation is preserved. This means that in the "receiving" sphere the WGM orbit is also tilted by the same angle in the opposite direction and that the direction of circulation of WGM induced in the "receiving" sphere is opposite to that in the first sphere. Since the coupling strength is directly related to the spatial overlap of two WGMs, moving the taper away from the equatorial plane results in increased gap between the modes that, in turn, results in their reduced coupling, as shown in Fig. 5(a).

This behavior can be modeled in simplified 2-D geometry by introducing a variable gap between two circular resonators. Such model adequately describes the underlying physical effect, but it does not take into account numerical solution in the 3-D case. In a simplified way depicted in Fig. 5(a) it can be imagined due to excitation of the real 3-D experimental geometry. To mimic our experiments we used resonators of 25 $\mu$m diameter with index of 1.57, medium index of 1.32, and a strip-waveguide of 1.5 $\mu$m width with effective refractive index of 1.45.

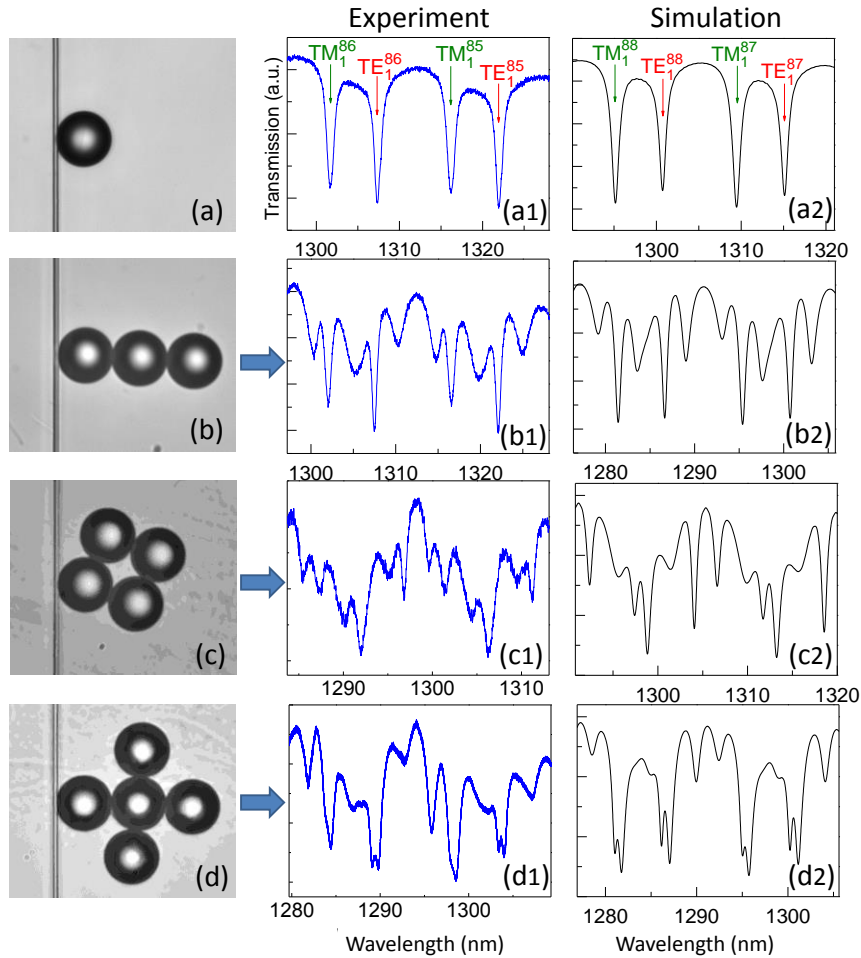

Figure 6. (a-d) Microscope images for various photonic molecules assembled with size-matched polystyrene microspheres of 25 $\mu$m mean diameter side-coupled to a tapered fiber with 1.5 $\mu$m waist diameter in water. (a1-d1) Measured and (a2-d2) simulated fiber transmission spectra for corresponding molecular configurations.

The simulation configuration is sketched in Fig. 5(c). Two touching circular resonators were placed in contact with the waveguide where femtosecond pulse with TE or TM polarization was launched in the strip-waveguide and the broadband transmission spectrum was recorded. The gap $g$ between two spheres was increased in 0.1 $\mu$m intervals. Several transmission spectra showing the mode splitting similar to that in the experimental spectra are presented in Fig. 5(d). Coupling constants can be derived in this method based on the mode splitting, which matches well with the corresponding experimental results presented in Fig. 5(b). The modal numbers for uncoupled WGM ($TE_{q=1}^{l=85}$) were determined by counting the number of E-field maxima along the radius and circumference of the circular resonator for much longer pulses under resonance with WGM. The $q$ and $l$-numbers in our 2-D model were found to be in a reasonable agreement with such numbers ($TE_{q=1}^{l=84}$) determined by fitting the WGM peak positions in the experimental spectra in Fig. 5(b) based on Mie scattering formalism. The comparison with the experiment in Fig. 6, supports our model based on the assumption that the effective gap created by moving the fiber away from the equatorial plane can be considered by analogy with the variation of the separation between two circular resonators.

It should be noted that our method of controlling the coupling constant is much simpler in practice compared to moving the resonators which has been used in previous studies [39, 40, 56]. Our method is especially convenient when changing the coupling constant is required simultaneously for many spheres forming more complicated structures. Such structures can be fabricated by various techniques including massively parallel manipulation with optoelectronic tweezers [57], directed self-assembly [58-61] or assembly of microspheres by air suction through array of micro holes [62]. The choice of the optimal technology is determined by a specific application. However, the proposed method allows tuning of coupling constant among all spheres by controlling only the position of the tapered fiber.

More complicated configurations were also assembled using pre-sorted spheres with almost identical resonant properties, including 3-sphere chain, planar quadrumer and 5-sphere cross, as shown in Fig. 6. First column (Figs. 6(a-d)) shows microscope images for the molecules assembled and brought in contact with the tapered fiber at their equatorial planes. The taper's contact positions were located by carefully adjusting the height while monitoring the spectra to find the largest mode splitting. The substrate's surface roughness in Figs. 6(c,d) is due to a thin layer (<5 $\mu$m) of epoxy used to glue the microspheres to the substrate. Due to the large size of spheres (25 $\mu$m diameter) there is no contamination near the equatorial plane that would affect the WGMs' $Q$-factors or coupling properties.

Experimental fiber transmission spectra with side coupling to a single sphere as well as to various photonic molecules are presented in Figs. 6(a1-d1). It should be noted that in order to reliably determine the spectral signature we need to record the transmission in a relatively wide spectral range. Since the excitation was provided by unpolarized white light source, both TE and TM WGMs are present in the experimental spectra, as indicated in Fig. 6(a1). The full transmission spectrum is a superposition of coupled WGM TE and coupled WGM TM modes in a molecule. Correspondingly, the simulation was performed with both TE and TM polarized light sources and the results were combined, as shown in Figs. 6(a2-d2).

Both experimental and simulated spectra contain periodically repeated spectral signatures formed by TE and TM polarized WGM-defined features near 1300 nm. Previously, attempts were made to observe the coupled modes spectrum in photonic molecules consisting of GaInAsP microdisks [17], however, the large size deviations (>1%) resulted in a merely satisfactory comparison to the calculated spectrum. Better uniformity was achieved in coupled microrings obtained by SIMOS technology [18], however some presence of disorder-induced spectral broadening effects and losses was still noticeable. In this work, using microspheres with size deviation of only 0.05% we demonstrated a very good agreement between experimentally measured and FDTD simulated spectra for both 1-D chain and 2-D planar molecules. The numbers of observed supermodes (dips) and their spectral positions match well in all cases. In some cases, like in Fig. 6(d1), the resolution of all WGM-related split components is not perfect. However, they are still visible (sometimes as shoulders) in both experimental and calculated spectra. These results demonstrate experimental feasibility of sphere sorting and photonic molecule assembling with high accuracy, and also provide strong support for our proposed concept of spectral signature identification discussed in Section 2.

## 6. Conclusions

In this work, we investigated the optical properties of photonic molecules consisting of coupled cavities with homogeneous WGM resonances both theoretically and experimentally. We proposed the concept of spectral signature associated with each molecular configuration as a relatively stable property which allows distinguishing between different molecules based on their spectra. The number of split supermodes and their spectral positions were studied based on the side-coupled fiber transmission spectra for 1-D and 2-D molecules formed by multiple atoms. We experimentally demonstrated the existence of such spectral signatures by assembling various molecules with pre-sorted polystyrene microspheres having 0.05% size deviation in an aqueous environment and by measuring the broadband transmission with a tapered fiber in contact with spheres at their equatorial plane. Spectral signature can be used as a unique feature for recognizing spatial configuration of a given molecule. It can find applications in position sensing as well as counterfeit technology.

Spatial distribution of the supermodes was also studied in simulation. Interesting coupling and transport phenomena such as Bloch mode formation, non-circular field pattern formation, long-range coupling of WGMs confined in circular resonators through surrounding medium, and the features of resonant optical tunneling through coupled cavities were observed and discussed. Further fundamental study of phase matching and coupling properties may be required to give more rigorous explanations. Photonic molecules also showed $Q$-factor enhancement for some of the antibonding modes and the mini-band formed by a series of overlapped supermodes. Therefore the spectrum of the molecule can be engineered, providing additional freedom in design of lasing devices, narrow-line filters, delay lines and multi-wavelength sensors. The E-field enhancement in the medium surrounding certain molecular configurations provides larger detecting volume and higher sensitivity comparing to typical WGM-based sensors of a single cavity.

Experimentally, we also demonstrated an alternative method of tuning the coupling constant between coupled spheres. By changing the position of mode excitation, which was realized by controlling the height of tapered fiber, we obtained complete tunability of the coupling constant from maximal value to zero in two touching spheres without moving them. This method is especially useful when coupling constant needs to be tuned simultaneously among all spheres assembled on a planar substrate.

As recently demonstrated, WGMs resonant enhancement of optical forces can be used to develop techniques for sorting microspheres with uniform resonant properties with ~$1/Q$ precision on a massive scale, which is applicable in both liquid [29] and air [33] environments. These pre-selected microspheres can be used as building atoms to construct complicated 2-D and 3-D photonic molecules whose properties have not been well studied. The present work proved the experimental feasibility of assembling a few spheres in various configurations in an aqueous environment that still maintain strong coupling and a high $Q$-factor. Large scale array can be fabricated by a variety of techniques [57-61]. Liquid environment is critical for particle movement and biomedical sample survival, however, compact photonic devices based on coupled microresonators can also be developed in air using high-index ($n$>1.9) spheres which can possess $Q$~$10^4$ with sufficiently small sizes ($D$<5 $\mu$m) [38].

**Acknowledgements.** The authors gratefully acknowledge support from U.S. Army Research Office through Dr. J. T. Prater under Contract No. W911NF-09-1-0450 and DURIP W911NF-11-1-0406 and W911NF-12-1-0538. This work was also supported by Center for Metamaterials, an NSF I/U CRC, Award No. 1068050. Also, this work was sponsored by the Air Force Research Laboratory (AFRL/RYD, AFRL/RXC) through the AMMTIAC contract with Alion Science and Technology and the MCF II contract with UES, Inc.